\documentclass{llncs}
\usepackage{booktabs} 
\usepackage{algorithm}
\usepackage{algorithmic}
\usepackage{todonotes}
\usepackage{amssymb}
\usepackage{multirow}
\usepackage{url}
\usepackage{rotating}
\usepackage{refcount}
\usepackage{paralist}
\usepackage{mathtools}
\usepackage{array} 
\usepackage{mdwlist} 
\usepackage{breqn}

\usepackage[font=small,labelfont=bf]{caption}

\begin{document}
\title{Analyzing Social Book Reading Behavior on Goodreads and how it predicts Amazon Best Sellers}

\author{%
Suman Kalyan Maity\inst{1}\thanks{This research has been performed when all the researchers were at IIT Kharagpur, India} \and Abhishek Panigrahi\inst{2} \and Animesh Mukherjee\inst{3}%
\institute{
Kellogg School of Management \\ and Northwestern Institute on Complex Systems, \\Northwestern University, Evanston IL 60208, USA \\ 
\email{suman.maity@kellogg.northwestern.edu}
\and Microsoft Research India \\
Bengaluru, Karnataka 560001, India \\
\email{abhishekpanigrahi034@gmail.com}
\and Department of Computer Science and Engineering, \\
Indian Institute of Technology Kharagpur, WB 721302, India\\
\email{animeshm@cse.iitkgp.ernet.in}\\}}
\maketitle
\begin{abstract}
A book's success/popularity depends on various parameters - extrinsic and intrinsic. In this paper, we study how the book reading characteristics might influence the popularity of a book. Towards this objective, we perform a cross-platform study of Goodreads entities and attempt to establish the connection between various Goodreads entities and the popular books (``Amazon best sellers''). We analyze the collective reading behavior on Goodreads platform and quantify various characteristic features of the Goodreads entities to identify differences between these Amazon best sellers (ABS) and the other non-best selling books. We then develop a prediction model using the characteristic features to predict if a book shall become a best seller after one month (15 days) since its publication. On a balanced set, we are able to achieve a very high average accuracy of {\bf 88.72\%} ({\bf 85.66\%}) for the prediction where the other competitive class contains books which are randomly selected from the Goodreads dataset. Our method primarily based on features derived from user posts and genre related characteristic properties achieves an improvement of 16.4\% over the traditional popularity factors (ratings, reviews) based baseline methods. We also evaluate our model with two more competitive set of books a) that are both highly rated and have received a large number of reviews (but are not best sellers) (HRHR) and b) Goodreads Choice Awards Nominated books which are non-best sellers (GCAN). We are able to achieve quite good results with very high average accuracy of {\bf 87.1\%} and as well a high ROC for ABS vs GCAN. For ABS vs HRHR, our model yields a high average accuracy of {\bf 86.22\%}.
\end{abstract}

%
%

\section{Introduction}
\begin{quotation}
 \small \textit{``If one cannot enjoy reading a book over and over again, there is no use in reading it at all.''} \\
 --- Oscar Wilde
\end{quotation}
Analysis of reading habits has been an active area of research for quite long time~\cite{Baumer:2008,Baumer:2011,Follmer:2012,Nardi:2004,Raffle:2010}. While most of these research investigate blog reading behavior~\cite{Baumer:2008,Baumer:2011,Nardi:2004}, there have been some work that also discuss about interactive and connected book reading behavior~\cite{Follmer:2012,Raffle:2010}. Despite such active research, very little investigation has been done so far to understand the characteristics of social book reading sites and how the collective reading phenomena can even influence the online sales of books. In this work, we attempt to bridge this gap and analyze the various factors related to book reading on a popular platform -- Goodreads and apply this knowledge to distinguish Amazon best seller books from the rest. 

Goodreads is a popular social book-reading platform that allows book lovers to rate books, post and share reviews and connect with other readers. On Goodreads website, users can add books to their personal bookshelves\footnote{In Gooreads, a book shelf is a list where one can add or remove books to facilitate reading similar to real-life book shelf where one keep books} for reading, track the status of their readings and post a reading status, find which books their friends are reading and what their favorite authors are reading and writing and get personalized book recommendations. Goodreads also promotes social interactions among users; users can participate in discussions and take part in group activities on various topics and allow users to view his/her friends' shelves, read reviews and comment on friends' pages. 

Popularity of a book depends on various factors. They can be broadly classified into two groups - intrinsic or innate content factors and external factors. Intrinsic content factors mostly concern quality of books that include its interestingness, the novelty factor, the writing style, the engaging story-line etc., in general. However, these content and quality factors of books are very different for different genres. For example, a successful thriller requires a credible, big story-line, strong narrative thrust, different view points, complex twists and plots, escalating stakes and tensions, breakneck speed with occasional lulls\footnote{https://hunterswritings.com/2012/10/12/elements-of-the-psychological-thriller-mystery-suspense-andor-crime-fiction-genres/} whereas a popular romantic novel does not require complex twists and plots, tension or shock effect; what it requires are variety, demonstration of strong and healthy relationship, once-in-a-lifetime love\footnote{http://www.writersdigest.com/wp-content/uploads/Essential_Elements.pdf}, conflicts, sexual tension etc.~\cite{hall}. It is ,therefore, difficult to find common grounds for books belonging to various genres and to quantify those aspects. External factors driving books' popularity include the readers' reading behavior, social contexts, book critics' reviews etc. In this work, we try to quantify the external factors of books' popularity by analyzing the characteristics of the entities and the book-reading behavior as reflected on the Goodreads platform. Particularly, we are interested to understand whether the collective reading behavior on Goodreads can distinguish the Amazon best sellers from the rest of the books.\\

\textbf{Research objectives and contributions:}
We analyze a large dataset from Goodreads to understand various characteristic differences existing between the Amazon best selling books and the rest and make the following contributions.
\begin{itemize}
 \item We study the characteristics of Amazon best sellers in terms of various Goodreads entities - books, authors, shelves, genres and user status posts. We observe that, across various features extracted from these entities, the best sellers are significantly different from an equal-sized sample of books selected uniformly at random.
 \item We leverage upon the characteristic properties of these best sellers and propose a framework to predict whether a book will become a best seller in the long run or not, considering review and reading behaviors of the books for various time period of observation $t$ = 15 days, one month post-publication. For one month, we achieve average accuracy {\bf 88.87\%} with average precision and recall of {\bf 0.887} for 10-fold cross-validation on a balanced set. The results are very similar for the other observation window. One of the most non-trivial results is that our user status and genre based prediction framework yields much better performance than traditional popularity indicators of books like ratings, reviews ($\sim 16.4$\% improvement). We would like to stress here that this result has a very important implication -- \textit{the Amazon best seller books might not necessarily be qualified by high quality reviews or a high volume of ratings; however, a large majority of them have user status post patterns that strongly distinguish them from the rest of the books}.

\item Since the number of ABS books should, in reality, be far lower than the other set of randomly chosen Goodreads books, we also evaluate our model under class imbalance in the test data set and achieve as good result as the balanced one. In specific, even for as small observation period as 15 days, we achieve a weighted average accuracy of $\sim$ \textbf{86.67\%} with weighted average precision of $0.901$ and recall of $0.867$ and a very high area under the ROC curve.
\item We further show that the proposed features can also discriminate between the best sellers and two competing sets of books that are a) highly rated, have a large volume of reviews but are not best sellers (HRHR) and b) Goodreads Choice Awards Nominated (GCAN) but not best sellers. The average accuracy of prediction for ABS vs GCAN is as high as {\bf 87.1\%} while for ABS vs HRHR, the average accuracy is \textbf{86.22\%}.
\end{itemize}

We believe that this work\footnote{This research is an extension of our earlier published work~\cite{Maity:2017} at ASONAM '2017 and reporting a much more detailed analysis emphazing various aspects of social book reading in more detail and perform detailed comparison of the best sellers with other kind of competitors} is an important contribution to the current literature as it not only unfolds the collective reading behavior of a social book-reading platform through a rigorous measurement study but also establishes a strong link between two orthogonal channels -- Goodreads and Amazon. Such a linkage might be extremely beneficial in fostering business for both the organizations through novel cross-platform policy designs. \\

\textbf{Organization of the paper:} The remainder of the paper is organized as follows: in the next section, we
discuss the state-of-the-art. In section 3, we describe the method for dataset preparation. Section 4 is devoted to the analysis of various characteristic properties of the Goodreads entities for the Amazon best sellers. In section 5, we discuss the prediction framework built for predicting whether a book will be an Amazon best seller or not and evaluate our proposed model. In section 6, we provide a discussion concerning the Amazon best sellers and the set of books with high ratings and reviews. In section 7, we draw conclusions pointing to the key contributions of our work and discuss potential future directions.
\section{Related Works}
There are some research works on the success of novels/books. Some of the early works~\cite{ell,harvey,mcgann} provide quantitative insights to stylistic aspects in successful literature relying on the knowledge and insights of human experts on literature. Harvey~\cite{harvey} and Hall~\cite{hall} focus mainly on content of the best-selling novels and try to prepare the secret recipes of the successful novels/books. Yun~\cite{yun2011} study the success of motion pictures based mainly on external, non-textual characteristics. Ashok et al.~\cite{ashok} focus on writing styles of the novels and establish the connection between stylistic elements and the literary success of novels providing quantitative insights to them.

One of the related domain of understanding success of books is that of text readability. Some of the early works propose various readability metrics based on simple characteristics of text documents like sentence length, number of syllables per word etc., for example, FOG~\cite{fog}, SMOG~\cite{smog} and Flesh-Kincaid~\cite{kincaid} metrics. More advanced readability measures based on list of predetermined words are the Lexile measure~\cite{lexile}, the Fry Short Passage measure~\cite{fry} and the Revised Dale-Chall formula~\cite{chall}. Recently, there have been several works~\cite{louis,kate,schwarm,hell,collins,pitler} that focus on predicting and measuring readability of texts based on linguistic features. Collins-Thompson and Callan~\cite{collins} adopt simple language modeling approach using a modified Na{\"i}ve Bayes classifier to estimate reading difficulty of texts. Heilman et al~\cite{hell} and Schwarm and Ostendorf~\cite{schwarm} use syntactic features apart from language model features to estimate grade level of texts.
Pitler and Nenkova~\cite{pitler} predicts readability of texts from Wall Street Journal using lexical, syntactic and discourse features. Kate et al.~\cite{kate} propose a model using syntactic and language model features to predict readability of natural language documents irrespective of genres whereas Louis~\cite{louis} propose a novel text quality metric for readability that considers unique properties of different genres.

Apart from understanding the success of books, readability of documents, there have been studies to identify author styles~\cite{raghavan,feng,peng,escalante,stam,baayen}. Peng et al.~\cite{peng} build a character-level $n$-gram language model for authorship identification. Stamatatos et al.~\cite{stam} use a combination of word-level statistics and part-of-speech counts or $n$-grams for author attribution. Baayen et al.~\cite{baayen} suggest that the frequencies with which syntactic rewrite rules are put to use provide a better clue to authorship than word usage and thus can improve accuracy of authorship attribution.

Another spectrum of works have been done in pursuit of understanding social blog and book reading behavior. Rideout et al.~\cite{zero} study shared book reading behavior and show that reading (or being read to) remains a constant in most young children's lives. Nardi et al.~\cite{Nardi:2004} examine the social nature of blogging activity and demonstrate that blogs are quite unlike a personal diary. Baumer et al.~\cite{Baumer:2008} perform a qualitative study focusing on blog readers, their reading practices, their perceptions of blogs and bloggers. The blogging activity is found to be far more heterogeneous and multifaceted than previously suggested. In a subsequent paper, Baumer et al.~\cite{Baumer:2011} study blog readers, their interactions with bloggers and their impact on blogging focusing on political blogs. Follmer et al.~\cite{Follmer:2012} introduce an interactive book-reading platform `People in Books' using FlashCam technology. The system supports children and their long-distance family members to interact via a play where the children and their family members can assume various characters/roles in children's story books over distance. Raffle et al. ~\cite{Raffle:2010} design `Family Story Play', a book-reading system which supports book reading among grandparents and their grandchildren over distance. Family Story Play establishes the hypothesis that there is a synergy between young children's education (a rich shared reading experience) and communication with long-distance family. 

There have been several works on book recommendations and author ranking. Huang and Chen~\cite{chen2005link} analyze user-item interactions as graphs and employ link prediction method for making collaborative filtering based recommendation for books on a book sales dataset. Kamps~\cite{kamps2011impact} investigates the effectiveness of author rankings in a library catalog to improve book retrieval. Vaz et al.~\cite{vaz2012improving} propose a hybrid recommendation system combining two item-based collaborative filtering algorithms to predict books and authors that the user will like. Zhu and Wang~\cite{zhu2007book} apply improved algorithm through filtering basic item set or ignoring the transaction records that are useless for frequent items generated to mine association rules from circulation records in university library. Vaz et al.~\cite{vaz2012stylometric} explore the use of stylometric features in the book recommendation task. Yang et al.~\cite{yang2009artmap} presents framework of clustering and pattern extraction by using supervised ARTMAP neural network by formation of reference vectors to classify user profile patterns into classes of similar profiles forming the basis of recommendation of new books. Givon and Lavrenko~\cite{givon2009predicting} try to solve the ``cold-start'' problem (books with no tags) in book recommendations by proposing a probabilistic model for inferring the most probable tags from the text of the book. Zhou~\cite{zhou2010book} analyzes the problem of trust in social network and proposes a recommender system model based on social network trust. Pera and Ng~\cite{pera2013read} propose a recommender system tailored to K-12 (Kindergarten to $12^{th}$ grade) readers, which makes personalized suggestions on books that satisfy both the preferences and reading abilities of its users. In a subsequent work, Pera and Ng~\cite{pera2014automating} propose a personalized book recommender system that emulates the readers' advisory process offered at public/school libraries to recommend books that are similar in contents, topics, and literary elements of other books appealing to a reader, with the latter based on extracted appeal-term descriptions. In another work, Pera and Ng~\cite{pera2015analyzing} propose unsupervised book recommendation framework for very young children especially K-3 (Kindergarten to $3^{rd}$ grade) readers.

On Goodreads platform, there has been very little research till date. Dimitrov et al.~\cite{dim} study the behavioral differences of reviews in Amazon and Goodreads. Adam Worrall~\cite{adam} have done an analysis on message posted by users in LibraryThing and Goodreads. Thelwall and Kousha~\cite{thel} provide insights in Goodreads users' reading characteristics. Thelwal~\cite{thelwall2017book} shows the existence of gender differences in authorship within genres for English language books. Our work differs from these above in following ways. We study the characteristics of various entities on Goodreads and try to establish whether these factors can discriminate between Amazon best sellers from other books. We use both the external characteristics of a book as well as the content of the reviews in our study. To the best of our knowledge, we are the first who try to explicitly provide quantitative insights, based on collective reading habits, on the unstudied connection between the entities of a book-reading platform (Goodreads) and the success of a book (best-sellers on Amazon). 

\vspace{-1mm}
\section{Dataset Preparation} We obtain our Goodreads dataset through APIs and web-based crawls over a period of 9 months. This crawling exercise has resulted in the accumulation of a massive dataset spanning a period of around nine years. We first identify the unique genres from \url{https://www.goodreads.com/genres/list}. Note that genres in the Goodreads community are user defined. Next we collect unique books from the above list of genres and different information regarding these books are crawled via Goodreads APIs. Each book has information like the name of the author, the published year, number of ratings it received, average rating, number of reviews etc. In total, we could retrieve information of 558,563 books. We then find out the authors of these books and their information like number of distinct works, average rating, number of ratings, number of fans etc. In total, we have information of 332,253 authors. We separately collect the yearly Amazon best sellers\footnote{\url{http://www.amazon.com/gp/bestsellers/1995/books}} from 1995 to 2016 and their ISBNs and then re-crawl Goodreads if relevant information about some of them is not already present in the crawled dataset. For these books, we separately crawl upto 2000 reviews and ratings in chronological order. We also crawl relevant shelves information of those books.

\section{Characteristic behavior} In this section, we shall study the characteristic properties of various Goodreads entities for the Amazon best sellers and compare them with the rest of the books. We have 1468 Amazon best sellers in our dataset. To compare with the rest of the books, we choose random samples of books from the entire set of books minus the Amazon best sellers. We obtain ten such random samples and report averaged results for them. Goodreads has three primary entities - books, authors and users. Here, we attempt to discriminate the Amazon best sellers from the others by analyzing these Goodreads entities. 

\subsection*{Book ratings and reviews} 
\begin{figure*}[h]
\begin{center}
\includegraphics[width=1\columnwidth]{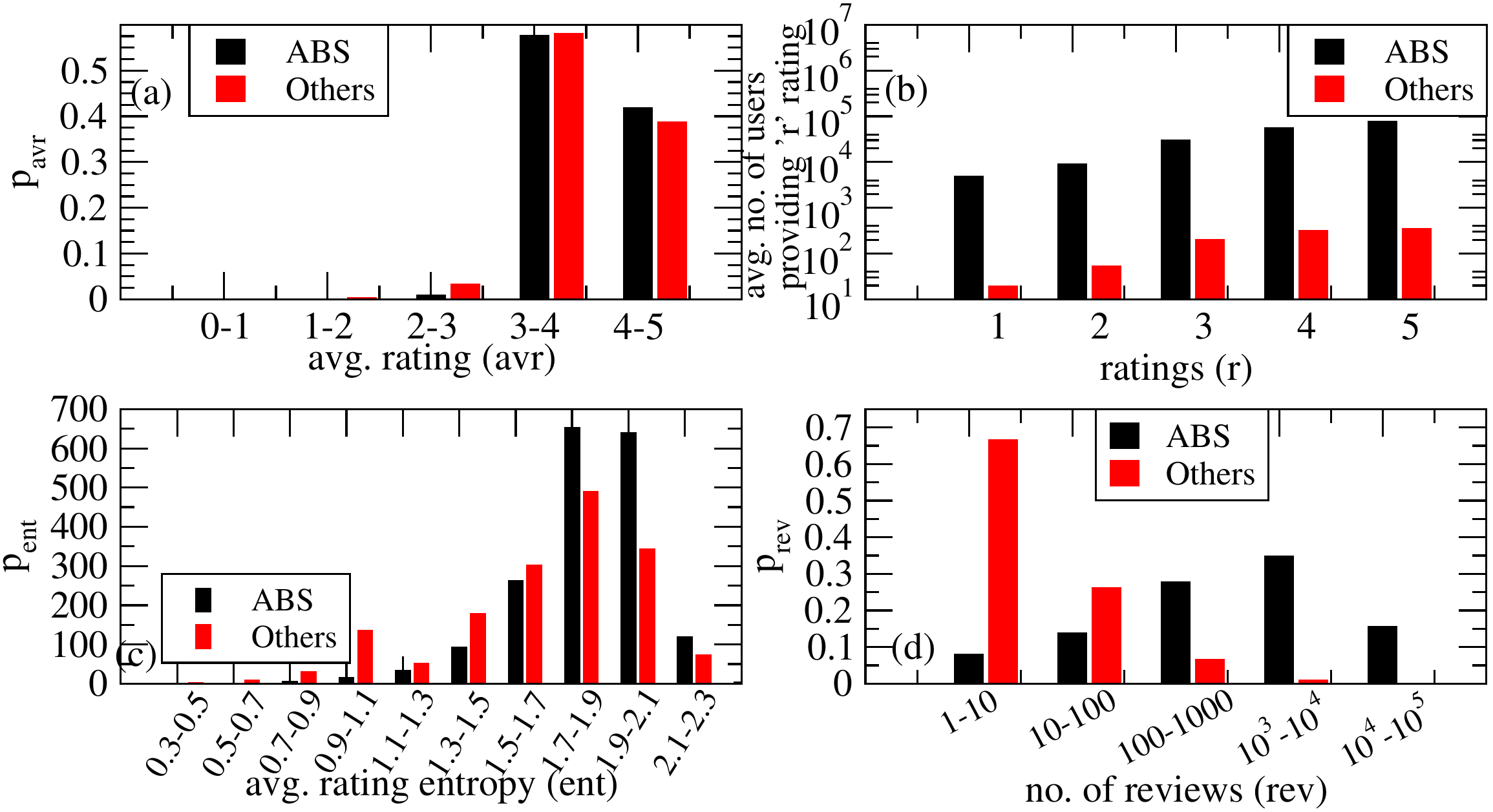}
\caption{\label{figbook} \small Book rating and reviews: distribution of a) average rating b) number of ratings c) average rating entropy d) number of reviews for Amazon best sellers (ABS) vs other books.}
\end{center}
\end{figure*}

Ratings and reviews received by a book are common factors that reflects aspects of collective reading and, in turn, governs the popularity of a book. We study the average rating distribution of the books in fig~\ref{figbook} (a). On average scale, there does not exist any significant difference in the ratings a best seller receives compared to the other books. In fig~\ref{figbook} (b), we compare the total number of ratings received -- Amazon Best Sellers receive more ratings than the other books, however, they receive both high and low ratings in larger proportions compared to the other books. We then calculate the rating entropy of the books which is a measure of diversity of ratings. Rating entropy ($rating_{entropy}$) of a book is formally defined as follows:
\begin{equation*}
 rating_{entropy}(b) = -\sum_{j \in {1..5}}p_j\times\log(p_j)
\end{equation*}
where $p_j$ is the probability of the book receiving a rating value $j$. Fig~\ref{figbook} (c) shows the distribution of the rating entropy of the books. The Amazon best sellers receive more diverse ratings compared to the other books. In fig~\ref{figbook} (d), we show the distribution of number of reviews. Amazon best selling books tend to get more reviews whereas majority of other books receive 10-100 reviews. 

%

\subsection*{Book genres and book shelves}
\noindent{\bf Genres}: Genres of books could be an important determinant of popularity. Not all of the genres attract equal number of readers. Some are very popular whereas some require very specific taste. In table~\ref{genre}, we observe the differences in various genres. $\sim57\%$ of the Amazon best sellers belong to the `Non-fiction' category, followed by the `Fiction' category constituting $\sim35\%$ of the books. Since one book can belong to multiple genres, the sum of all these fractions denoting the fraction of books belonging to that genre, do not sum up to 1. For books that are not best sellers (i.e., others), the rank order for fiction and non-fiction gets reversed. Another striking difference in genres is that `Self Help', `Reference', `Adult', `Business' etc. has higher contribution of books to the Amazon best selling books compared to the set of other books. As we shall see, the genre characteristics turn out to be one of the most distinguishing factors between the Amazon best sellers and the other books.
\begin{table*}[!t]
\begin{center}
\caption{\small Top 15 genres corresponding to each category of books.}
\label{genre}
\begin{tabular}{|p{3cm}|p{3cm}|p{3cm}|p{3cm}|}\hline
\multicolumn{4}{|c|}{Categories} \\ \hline
Amazon Best Sellers & Fraction of books & Others & Fraction of books  \\\hline
Non-fiction & 0.577 & Fiction & 0.221 \\ 
Fiction & 0.349 & Nonfiction & 0.202 \\ 
Self Help & 0.219 & Romance & 0.119 \\ 
Adult & 0.219 & Fantasy & 0.103 \\ 
Reference & 0.19 & Literature & 0.103 \\ 
Business & 0.185 & History & 0.091 \\ 
History & 0.179 & Cultural & 0.084 \\ 
Biography & 0.151 & Historical & 0.063 \\ 
Science & 0.143 & Children & 0.063 \\ 
Novels & 0.143 & Historical Fiction & 0.06 \\ 
Psychology & 0.141 & Mystery & 0.06 \\ 
Literature & 0.137 & Contemporary & 0.058 \\ 
Fantasy & 0.131 & Sequential Art & 0.054 \\ 
Contemporary & 0.128 & Science Fiction & 0.051 \\ 
Adult Fiction & 0.117 & Young Adult & 0.051 \\ \hline
\end{tabular}
\end{center}
\end{table*}

\begin{figure*}[h]
\begin{center}
\includegraphics[width=1\columnwidth, angle=0]{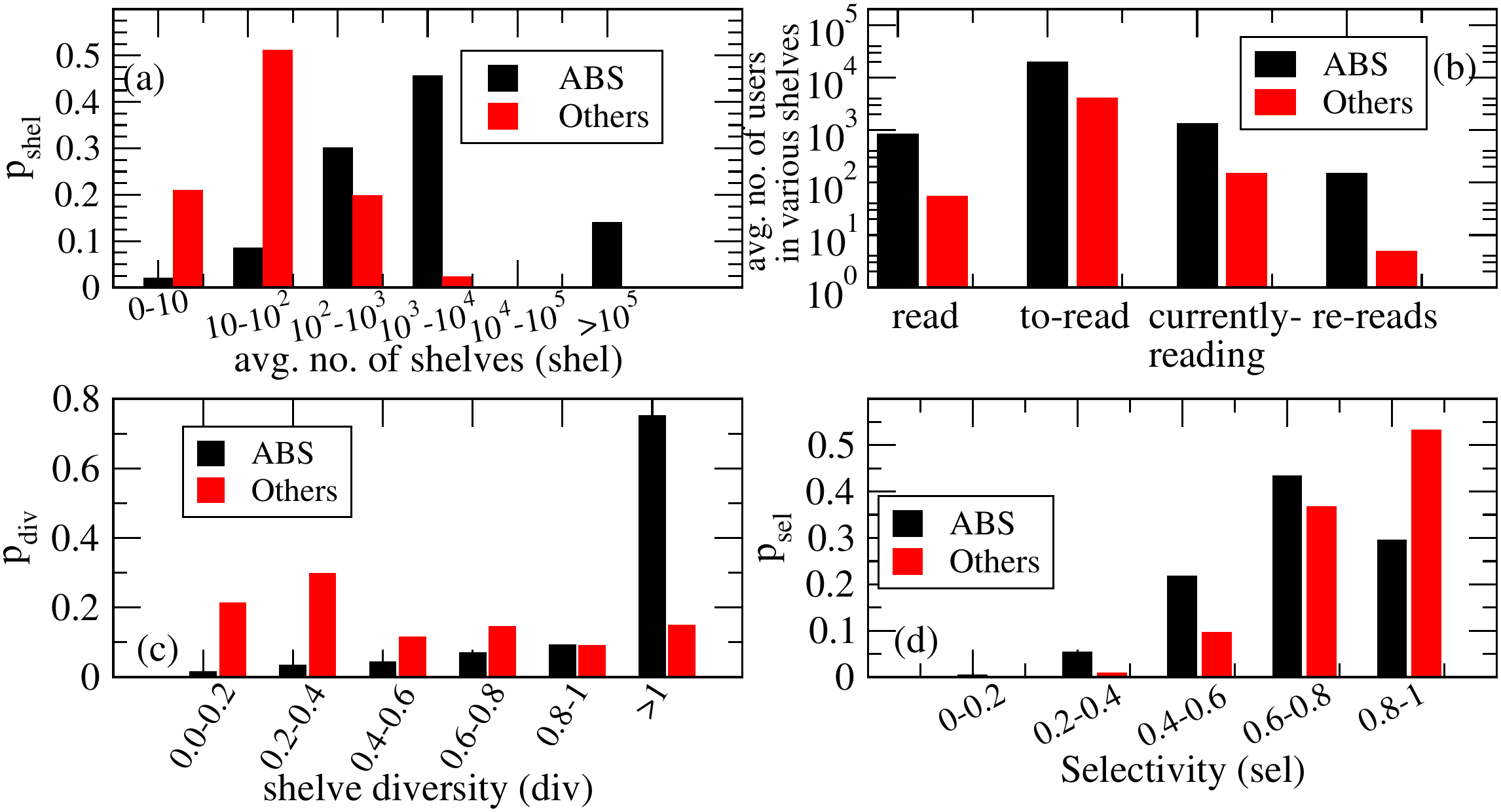}
\caption{\label{figshelves} \small Characteristic properties of book shelves: distribution of a) number of shelves b) average number of users tagging their books to `read', `to-read', `currently-reading' and `re-reads' shelves c) Shelf diversity d) Selectivity for ABS vs other books.}
\end{center}
\end{figure*}

\noindent{\bf Shelves}: To facilitate ease of book reading, Goodreads provides a unique feature of organizing books into various shelves. It provides three default self-explanatory bookshelves: `read', `currently-reading', `to-read' and also provides opportunity to the user to create customized shelves to categorize his/her books. We shall analyze these book shelves to identify if they are relevant for popularity/success of a book. Fig~\ref{figshelves} (a) shows the distribution of number of bookshelves a book is kept in. We observe that the Amazon best sellers are placed in a much larger number of book shelves (as large as $>10^5$) by the users compared to the rest of the books. We then concentrate on the content of four specific shelves and their variants - `read', `to-read', `currently-reading' and `re-reads' shelves. We observe that for the Amazon best sellers, all these shelves are more dense compared to the other books (see fig~\ref{figshelves} (b)).

\textbf{Shelf diversity:} Similar to rating entropy defined earlier, we calculate shelf diversity which quantifies the idea of how users put their books in various shelves. Formally, shelf diversity (ShelfDiv) can be defined as follows:
\begin{equation*}
 ShelfDiv(b) = -\sum_{j \in shelf_{set}}s_j\times\log(s_j)
\end{equation*}
where $s_j$ is the probability that the book belongs to the $j^{th}$ shelf in the set of book shelves.
Fig~\ref{figshelves} (c) shows that the Amazon best sellers have higher shelf diversity. $\sim75\%$ of the best sellers have diversity score of $> 1$. 

\textbf{Selectivity:} We define a selectivity metric which tells us how users are selective in putting the books in their shelves. More formally, $k$-shelf selectivity of a book is defined as the fraction of users being covered if one selects only top-$k$ shelves used for keeping the book. In fig~\ref{figshelves} (d), we show the distribution of {\em selectivity}. Here, we show $3$-shelf selectivity of the books. It is observed that the readers of the Amazon best sellers are less selective compared to the readers of the other Goodreads books. 

\subsection*{Goodreads users' status posts} 
\begin{figure*}[h]
\begin{center}
\includegraphics[width=1\columnwidth, scale=0.5, angle=0]{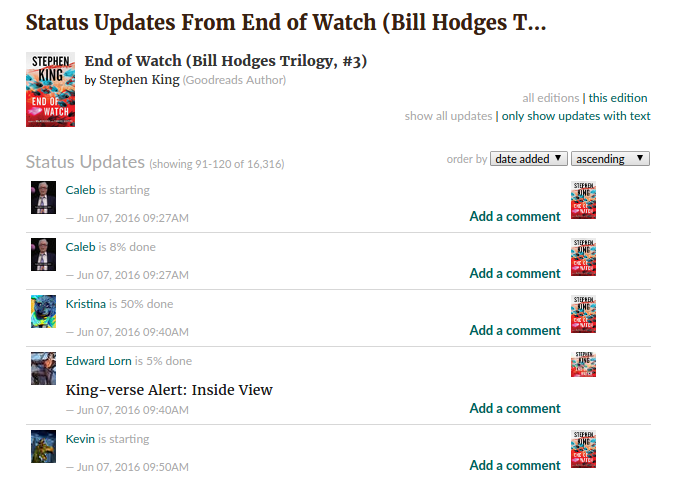}
\caption{\label{exstatus} Status page for ``End of Watch (Bill Hodges Trilogy, \#3) by Stephen King''}
\end{center}
\end{figure*}

While reading, a Goodreads user can post status updates about reading the book. For example, one can post how much of the book has been read, which page he/she is in, also can comment about the book so far read etc. (see fig~\ref{exstatus}) We separately crawl the first 2000 user status posts for each book in our dataset. These book reading status postings for a book also drive its popularity. We attempt to differentiate the Amazon best sellers with the other Goodreads books through the reading status postings. 

\begin{figure*}[h]
\begin{center}
\includegraphics[width=1\columnwidth, angle=0]{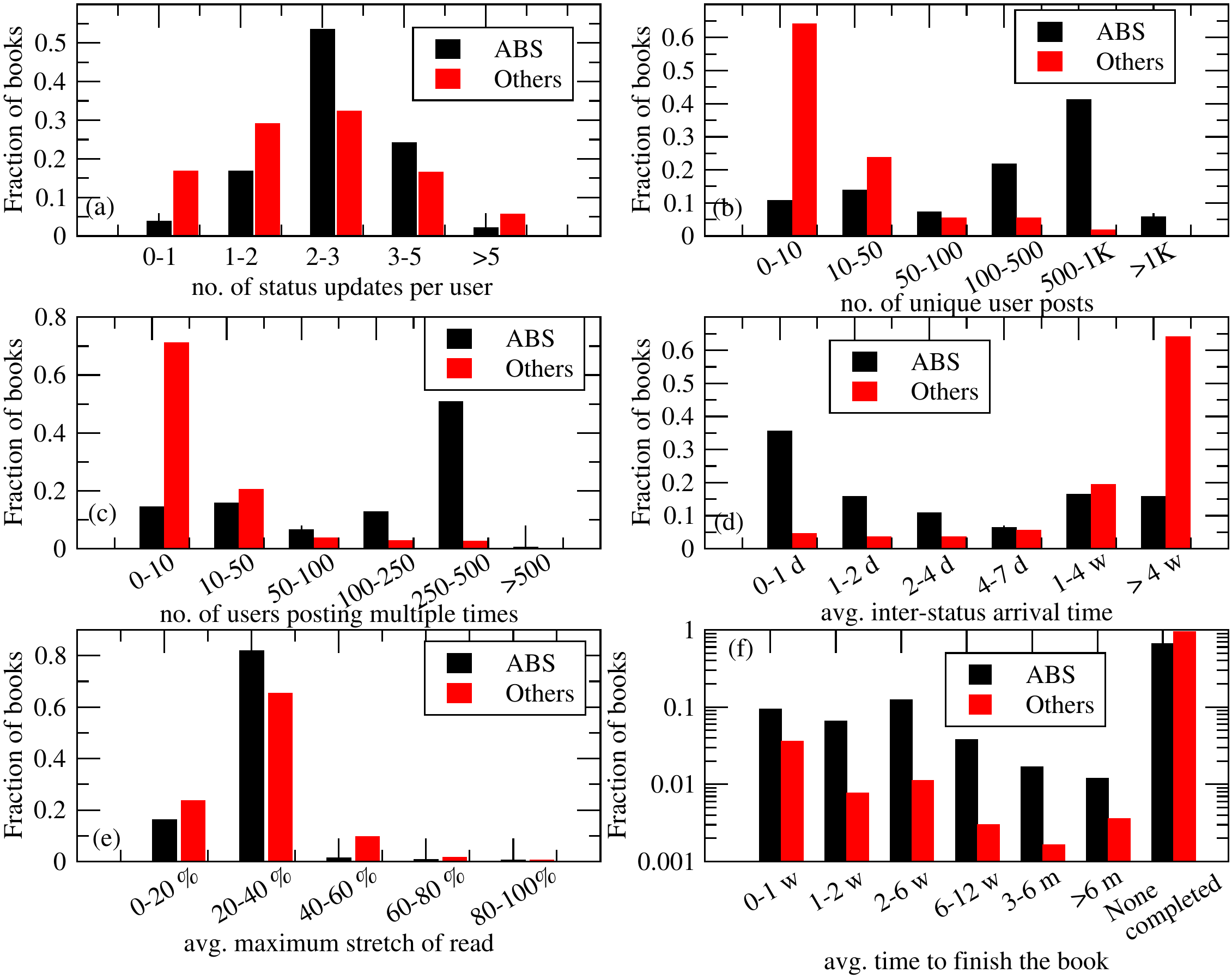}
\caption{\label{figstatus} \small Characteristic properties of Goodreads users' status posts: distribution of a) number of status updates per user  b) number of unique users updating status c) number of users updating multiple times d) inter-status arrival time e) average maximum stretch of reading f) average time to finish reading for ABS vs other books.}
\end{center}
\end{figure*}

In fig~\ref{figstatus} (a), we show the distribution of the number of status update posts per users. The results show that while reading the Amazon best seller books, readers tend to post status updates more often as compared to the readers of other books. Fig~\ref{figstatus} (b) presents the distribution of unique readers posting status updates. The Amazon best sellers engage more readers in posting status updates compared to other Goodreads books. Also, the Amazon best sellers engage better the same readers in posting multiple status updates compared to other Goodreads books (see fig~\ref{figstatus} (c)). We also study the distribution of average inter-status arrival time (see fig~\ref{figstatus} (d)) which shows that readers of Amazon best sellers post status updates more frequently (for more than $35\%$ books, average inter-status arrival time is less than a day) than the readers of other Goodreads books. Fig~\ref{figstatus} (e) shows the distribution of average maximum concentrated reading efforts at a stretch (in terms of percentage read). Readers of $\sim80\%$ of the Amazon best selling books show maximum percentage stretch of read as 20-40\%. Though, for the other Goodreads books also, the fraction is largest for the same zone, the relative number of books in this stretch is lesser than that of the best sellers. In fig~\ref{figstatus} (f), we show the distribution of average time to finish book reading. We observe that for $\sim63\%$ of the Amazon best sellers there are no readers who have completed reading the whole book whereas for the other books, this number is quite high ($\sim94\%$). Among the books where one of the readers have at least finished reading the book, the fraction of books for Amazon best sellers are higher compared to the other books in all the time buckets. Note that features extracted from the status posts of users strongly discriminate the Amazon best sellers from the rest of the books (see section~\ref{predict}).

\subsection*{Author characteristics}
\begin{figure*}[h]
\begin{center}
\includegraphics[width=1\columnwidth,angle=0]{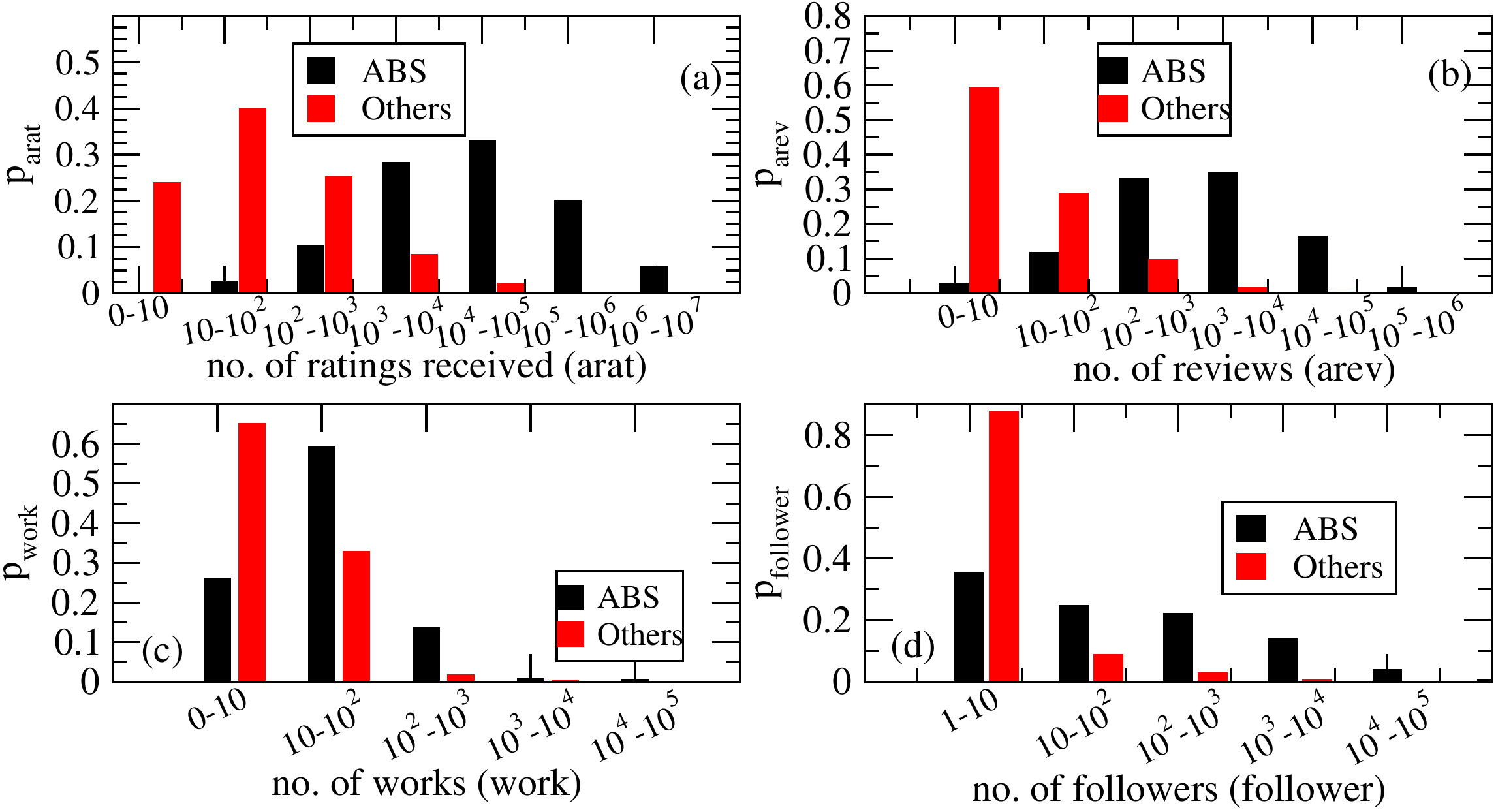}
\caption{\label{figauthor} \small Characteristic properties of the authors of the book: distribution of a) number of ratings received by the author b) number of reviews received c) number of literary works d) number of followers for ABS vs other books.}
\end{center}
\end{figure*}

Authors play an important role in driving the popularity of a book. Already successful or highly reputed authors have higher chance of having their books more popular than a novice writer. In fig~\ref{figauthor}, we show the distribution of various author features of the Amazon best sellers and compare them with that of the other set of Goodreads books selected uniformly at random. In fig~\ref{figauthor} (a), we show the distribution of number of ratings received by the authors. We observe that the authors of Amazon best sellers receive more ratings (more than $30\%$ of the authors of the best selling books receive $10^4-10^5$ ratings from the book readers; also a significant number of authors receive more than million ratings) compared to the authors of the other books ($\sim40\%$ of these authors receive $10-100$ ratings from the book readers). Fig~\ref{figauthor} (b) presents the reviews that the books of the authors receive. Likewise ratings, the number of reviews received also follow a similar distribution. The authors of the Amazon best sellers tend to receive higher number of reviews with $\sim51\%$ of them receiving more than 1000 reviews whereas the authors of the other Goodreads books receive less number of reviews (more than 60\% of these authors receive less than 10 reviews).

In fig~\ref{figauthor} (c), the distribution of number of works is shown. $\sim60\%$ of the authors of the Amazon best sellers pen around $10-100$ literary works whereas more than $60\%$ of the authors of the other Goodreads books on average pen less than 10 books. There are also some authors (mostly publishing houses) who publish more than $10^4$ works and this number is much higher in case of the Amazon best sellers.  Fig~\ref{figauthor} (d) demonstrates the characteristic differences in follower/fans distribution of the authors of various categories of books. The authors of the Amazon best sellers receive large number of fan following on Goodreads -- whereas $88\%$ of the authors of the other books on Goodreads have less than 10 fans, a large fraction of the Amazon best sellers' authors have more than 10,000 fans/followers. 

\section{Will a book become an Amazon best seller?}\label{predict}
From the discussions in the previous sections, it is evident that there exist differences among various Goodreads entities for Amazon best sellers and the other Goodreads books. In this section, we attempt to build a prediction framework that can early predict whether a book will be an Amazon best sellers or not. We shall try to leverage upon the various characteristic properties of the Goodreads entities that we have studied earlier. Note that many of these features for example number of followers of an author, the behavior of shelves etc. are temporal in nature; however, due to limitations on data crawl imposed by Goodreads we do not possess the temporal profile of all these features. Therefore, we shall attempt to build the predictors using minimal external and content features and will evaluate our framework against standard evaluation metrics like accuracy, precision, recall etc. Our goal is to predict whether a book will be an Amazon best seller or not just by observing data upto various time periods from Goodreads starting from the date of publication of the book ($t$ = 15 days, 1 month). As Goodreads was launched in 2007, we consider only those Amazon best seller that are published on or after 2007. To compare against these set of books, we consider a same sized random sample of books from Goodreads, none of which ever became an Amazon best seller and all of which have been published after 2007. In total, we have $\sim380$ books in each class. For the task of prediction, we consider the following set of minimal features:
\begin{itemize}
 
 \item \textbf{Novelty of the book:} Novelty of a book is a key for its acceptability/success in the readers' circle. For each book in Goodreads, a short summary about the book is provided. We separately crawl this `About' information of the book (say, the document containing the summary of the book be $A$) and of all the other books that are published before this book in question (say, the concatenated summary of all those books be $B$). We then extract keywords\footnote{\url{https://github.com/aneesha/RAKE}}~\cite{rake} from documents $A$ and $B$ respectively ($Keywords_A$ and $Keywords_B$) and design a metric of keyword overlap as follows:\\
 \begin{equation*}
 \small
Overlap(A, B) = \frac{|Keywords_A \cap Keywords_B |}{min (|Keywords_A|, |Keywords_B|)}
\end{equation*}
Higher the keyword overlap, lower is the novelty score. We use this as a feature for our model. \\
We also define another novelty feature which is measured as the Kullback-Leibler (KL) divergence between the unigram language model inferred from the document $A$ containing all the words except the stop words from the book's summary for the $i^{th}$ book and the background language model from document $B$ and formally defined this as follows:
\begin{equation*}
  KLDiv (i)= -\sum_{w \in A}p(w|A)\times\log\frac{p(w|A)}{p(w|B)}
\end{equation*}
Higher the divergence value, higher is the novelty of the book. Once again this is used a feature in the prediction model.
 \item \textbf{Genres} - As mentioned earlier, genres of a book is an important feature in deciding the destiny of it. In this prediction model, we consider top 15 genres for Amazon best sellers and other books as shown in table~\ref{genre}. In total we have 24 genres - fantasy, fiction, nonfiction, children, romance etc. We use each of them as a binary feature.
\end{itemize}
\textbf{Goodreads book reading status posts:} We compute several features out these book reading posts. The list is as follows:
\begin{itemize}
 \item number of status updates of the readers (mean, min, max and variance),
 \item number of unique readers posting status updates,
 \item number of readers posting status updates more than once (twice/thrice/five times),
 \item inter-status arrival time (mean, min, max and variance),
 \item maximum percentage stretch of read (mean, min, max and variance). We also use the maximum stretch of read in terms of time.
 \item rate of reading of the readers (mean, min, max and variance),
  \item fastest rate of reading (mean, min, max and variance),
 \item time taken to finish reading the book (mean, min, max and variance),
 \item average positive and negative sentiments from the status posts.
\end{itemize}

\textbf{Baseline features:}
Ratings and reviews of book are the most common indicators of popularity and in this paper, we shall consider features extracted from them as baselines and compare them with the more non-trivial features related to reading behavior of users for the task of prediction. The ratings and reviews related features are described below
\begin{itemize}
\item average number of rating
 \item number of 1-star ratings, 2-star ratings, 3-star ratings, 4-star ratings, 5-star ratings
 \item rating entropy
 \item number of reviews received
 \item \textbf{Sentiment of the reviews:} For each book, we concatenate all the reviews in one month and find out the fraction of positive sentiment words (positive sentiment score) and the fraction of negative sentiment words (negative sentiment score) by using MPQA sentiment lexicon~\cite{mpqa}. We use these two sentiment scores as two separate features.
 \item \textbf{Cognitive dimension of the reviews:} There could be differences in the cognitive dimension (linguistic and psychological) for the two category of books. To quantify this, we consider Linguistic Inquiry and Word Count (LIWC) software~\cite{liwc}. LIWC takes a text document as input and outputs a score for the input over all the categories based on the writing style and psychometric properties of the document.
 \end{itemize}

\subsection{Performance of the prediction model}
In this subsection, we shall discuss the performance of our prediction model. We use 10-fold cross-validation technique and use SVM and logistic regression classifier~\cite{weka}. For the prediction task, we consider $t$ time periods ($t$) - 15 days, 1 month from the publication date. We compute all the feature values from the data available only within the time period $t$ from the publication date. We ensure that all the books that we select in both the classes are published after 2007 since Goodreads was launched in 2007. Table~\ref{tabeval} shows the various classification techniques we employ and the evaluation results. The classifiers yield very similar classification performance with Logistic Regression performing little better; with logistic regression classifier, we obtain average accuracy of 88.72\% with average precision and recall of 0.887 each and the average area under the ROC curve as 0.925 for $t = 1$ month on a balanced dataset with 10-fold cross-validation method. Note that the classification results for other time period also give very similar results. The user status and genre based features are most prominent ones and significantly outperforms the ratings and review feature based baselines. For $t$ = 1 month, our method yields 16.4\% improvement over the best performing baseline (for $t$ = 15 days, we also yield similar improvement) suggesting that user's status on Goodreads are very important indicators of popularity and are, in fact, much better indicators than reviews or ratings. In other words, this shows that all Amazon best seller books might not necessarily have high quality reviews or a high volume of ratings; however, a large majority of them have user status post patterns very different from the other set of books.

Since in real-life, the proportion of the Amazon best sellers is far lower than the other types of books, we also consider testing our model on an unbalanced test set. Here, the training and test sample sets are taken in 3:1 ratio. In training set, both the class samples are taken in equal proportion (to guarantee fair learning) whereas in test sample the Amazon best sellers and the other books are taken in 1:9 ratio. We then train our classifiers on the balanced training set and test on the unbalanced one. We report the weighted average values for all the metrics in table~\ref{tabeval}. For an observation period of even as small as 15 days, we achieve weighted average accuracy of $\sim86.67\%$ with weighted average precision of $0.901$ and recall of $0.876$. Note that compared to the balanced set, the performance is slightly better. The weighted average ROC area under the curve is quite high ($0.963$ on weighted averaging - but same value is found for the individual classes also). 
\begin{table}[!ht]
\caption{ \small Evaluation results with comparison with baselines (baseline1 - ratings, baseline2 - reviews)}
\label{tabeval}
\begin{tabular}{ |p{1.5cm}|p{2.3cm}|p{1.5cm}|p{1.5cm}|p{1.5cm}|p{1.5cm}|p{1.5cm}| }
\hline
 $t$ & Method & Accuracy &Precision & Recall & F-Score & ROC Area \\ \hline
 \multirow{4}{1.5cm}{15 days}& LR &  \textbf{85.66\%}& 0.857   &  0.857  &  0.857 & \textbf{0.917} \\\cline{2-7}
  & SVM  & 85.3\% &0.853   &  0.853  &   0.853    &  0.853 \\ \cline{2-7}
  & Baseline1 - LR & 76.7\% & 0.774    &  0.767  &   0.766  &    0.826 \\\cline{2-7}
  & Baseline2 - LR & 75.98\% & 0.76   &   0.76    &  0.76     &  0.829 \\\hline
 \multirow{4}{1.5cm}{1 month}&LR & \textbf{88.72\%}& 0.888    & 0.887  &   0.887 & \textbf{0.925} \\\cline{2-7}
  & SVM  & 88.71\% &0.888   &  0.887  &   0.887   &   0.887\\ \cline{2-7}
  & Baseline1 - LR & 76.22\% & 0.766  &   0.762 &    0.762 &     0.808 \\\cline{2-7}
  & Baseline2 - LR & 75.91\% & 0.76    &  0.759     & 0.759   &   0.812 \\\hline
  \multirow{4}{1.5cm}{Unbalanc-ed testset (t = 15 d)}& LR & \textbf{86.67\%}& 0.901 &  0.867  &   0.876 &  \textbf{0.963}\\\cline{2-7}
  & SVM  & 86.67\% & 0.924 & 0.867  & 0.879   & 0.919 \\ \cline{2-7}
  & Baseline1 - LR & 75.56\% & 0.897  &   0.756 &    0.784  &    0.851 \\\cline{2-7}
  & Baseline2 - LR & 75.5\% &  0.839    &  0.756 &    0.78   &   0.78  \\\hline
  \end{tabular}
  \end{table}

%

\subsection*{Discriminative power of the features}
Here, we shall discuss about the importance of the individual features (i.e., the discriminative power of the individual features). In order to determine the discriminative power of each feature, we compute the chi-square ($\chi^2$) value and the information gain. Table~\ref{tab:featurerank} shows the order of all features based on the $\chi^2$ value, where larger the value, higher is the discriminative power. The ranks of the features are very similar when ranked by information gain (Kullback-Leibler divergence). The most prominent features on individual level are the user status features. There are several user status features that come in list of top 15. Among those status features, the most discriminative ones are mean reading rate, number of status posts, fastest rate of reading, inter-status post time etc.

\begin{table}[!htb]
\centering
\caption{\small Features and their discriminative power.}
\label{tab:featurerank}
 \begin{tabular}{ |p{1.5cm}|p{2cm}|p{7cm}| }
\hline
$\chi^2$ Value & Rank  & Feature  \\ \hline
194.0324& 1 & mean reading rate \\
121.7847  & 2 & number of readers posting status \\
109.8861   &  3 & fastest rate of reading (min) \\
 96.0871   &  4 & inter-status post time (min) \\
 92.1124  &  5 & min. reading rate \\
 88.2171   &  6 & number of readers posting status updates twice \\
 82.5114  &  7  & variance of reading rate \\
 75.9434  &  8 & maximum percentage stretch of read (min) \\
 75.0442   &  9 & number of readers posting status updates thrice\\
 74.0946   &  10 & inter-status post time (mean) \\
 71.7962  &  11 & time taken to finish the book (max) \\
 71.7024  &  12 & maximum percentage stretch of read w.r.t time (max) \\
 69.5682  &  13 & number of readers posting status updates five times \\
 63.3696   & 14 & genre (Romance) \\
 56.759   &  15 & time taken to finish the book (variance) \\ \hline
\end{tabular}
\end{table}
\section{Close competitors}
In this section, we shall study the characteristic behavior of the Amazon best sellers, contrasting them with two sets of close competitors described below. 

\textbf{HRHR}: We have observed in the earlier section that Amazon best sellers tend to have high ratings and reviews. Therefore, we retrieve a set of books from the Goodreads dataset that are published on or after 1995 and have received average rating of 4 or more and received at least 900 reviews. Note that the parameters of ratings and reviews are chosen in such a way so that we get a comparable number of books as that of the Amazon best sellers. We have termed these highly rated and reviewed non-best selling competitor books as `HRHR'. HRHR actually constitutes a weaker contrast class compared to the random set of books.

\textbf{Goodreads choice awards nominee}: Goodreads readers choice award\footnote{\url{https://www.goodreads.com/choiceawards/}} is an annual event. This was first launched in 2009. From then on, Goodreads users can take part in this award to nominate as well as to vote their nominations. There are 20 categories of awards and in each category 20 nominations are made. We have considered the non-best selling Goodreads choice awards' nominees (GCAN) from 2009 to 2015 as another competitive set to compare with the Amazon best sellers.

\subsection*{Comparisons}
\noindent\textbf{Genre characteristics}: First we study the competitors from the point of view of genres. In table~\ref{compgenre}, we show top 10 genres from each of these categories. We observe that there exist significant differences in genre distribution. Amazon best sellers books appear most in ``Non-fiction'' category whereas $\sim74\%$ of the HRHR books as well as $\sim87\%$ of the GCAN books appear in the ``Fiction'' category. Also, these competitor books belong to ``Adult'', ``Fantasy'', ``Contemporary'', ``Romance'' and other ``Fiction'' category in large proportions. Another interesting observation is that quite a large proportion of these competitor books ($\sim30\%$ for HRHR books and $\sim75\%$ for GCAN books) belong to the ``Adult'' category compared to a much smaller proportion ($\sim22\%$) for the Amazon best sellers. Similarly, a large proportion of competitor books ($\sim33\%$ for HRHR books and $\sim50\%$ for GCAN books) belong to ``Young Adult'' category.
\begin{table}[!t]
\begin{center}
\caption{\small Top 15 genres corresponding to each category of books.}
\label{compgenre}
\begin{tabular}{|p{2.5cm}|p{1.2cm}|p{2.5cm}|p{1.2cm}|p{2.5cm}|p{1.2cm}|}\hline
\multicolumn{6}{|c|}{Categories} \\ \hline
Amazon Best Sellers & Fraction & HRHR & Fraction& GCAN & Fraction  \\\hline
Non-fiction & 0.577 & Fiction & 0.737 & Fiction&0.875 \\
Fiction & 0.349 & Fantasy & 0.464& Adult&0.75 \\
Self Help & 0.219 & Romance & 0.388& Contemporary&0.625 \\
Adult & 0.219 & Young Adult & 0.332& Historical Fiction&0.625 \\
Reference & 0.190 & Adult & 0.298& Mystery&0.625 \\
Business & 0.185 & Mystery & 0.212& Adult Fiction&0.625 \\
History & 0.179 & Contemporary & 0.207& Young Adult&0.5 \\
Biography & 0.151 & Adventure & 0.186& Fantasy&0.5 \\
Science & 0.143 & Historical Fiction & 0.159& Autobiography&0.375 \\
Novels & 0.143 & Literature & 0.159& Nonfiction&0.375 \\ 
Psychology & 0.141 & Historical & 0.156& Biography&0.375 \\
Literature & 0.137 & Science Fiction & 0.155& Biography Memoir&0.375 \\
Fantasy & 0.131 & Paranormal & 0.145& Historical&0.375 \\
Contemporary & 0.128 & Classics & 0.144& Thriller&0.375 \\
Adult Fiction & 0.117 & Childrens & 0.137 & Science Fiction&0.375 \\\hline
\end{tabular}
\end{center}
\end{table}

\begin{figure*}[h]
\begin{center}
\includegraphics[width=1\columnwidth,angle=0]{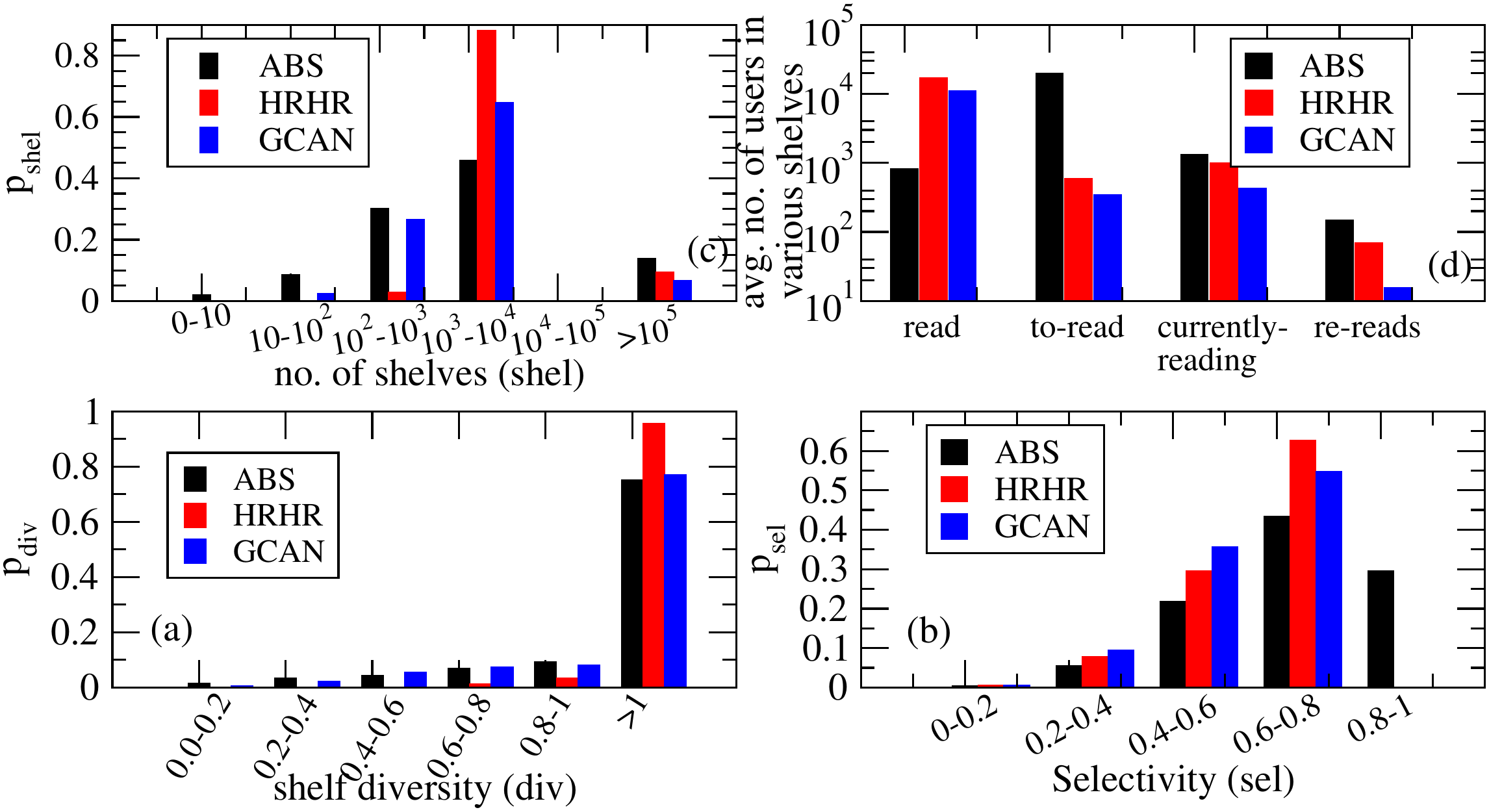}
\caption{\label{figcompshelves} \small Characteristic properties of book shelves: a) distribution of number of shelves b) average number of users tagging their books to `read', `to-read', `currently-reading' and `re-reads' shelves c) Shelf diversity d) Selectivity for ABS vs HRHR and GCAN books.}
\end{center}
\end{figure*}

\noindent\textbf{Book shelves characteristics}: We then attempt to differentiate the Amazon best sellers from the competitor books in terms of their book shelf properties. In fig~\ref{figcompshelves}, we show distribution of various shelf properties for Amazon best sellers vis-a-vis the competitor books. An interesting observation is that most of the competitor books ($\sim88\%$ for HRHR and $\sim65\%$ for GCAN) are parts of a large number of shelves ($10^3-10^4$) while this proportion is significantly smaller for Amazon best sellers ($\sim40\%$) (see fig~\ref{figcompshelves} (a)). Also these close competitor books have been `read' by more than 10 times more readers than the Amazon best sellers; on the other hand, for the Amazon best sellers more users (more than 10 times) have put them into `to-read' shelf compared to these competitors. Therefore, the competitor books are already read by a significant number of users on average whereas the Amazon best sellers are successful in engaging more users to still read them (see fig~\ref{figcompshelves} (b)). In terms of re-reading behavior, more readers re-read the Amazon best sellers than the competitor books. The competitor books are put in a higher variety of different shelves by the users leading to higher shelf diversity compared to the Amazon best sellers (see fig~\ref{figcompshelves} (c)) whereas the Amazon best sellers readers are more selective in placing the books in the shelves compared to the competitor books (see fig~\ref{figcompshelves} (d)).

\begin{figure*}[h]
\begin{center}
\includegraphics[width=1\columnwidth, angle=0]{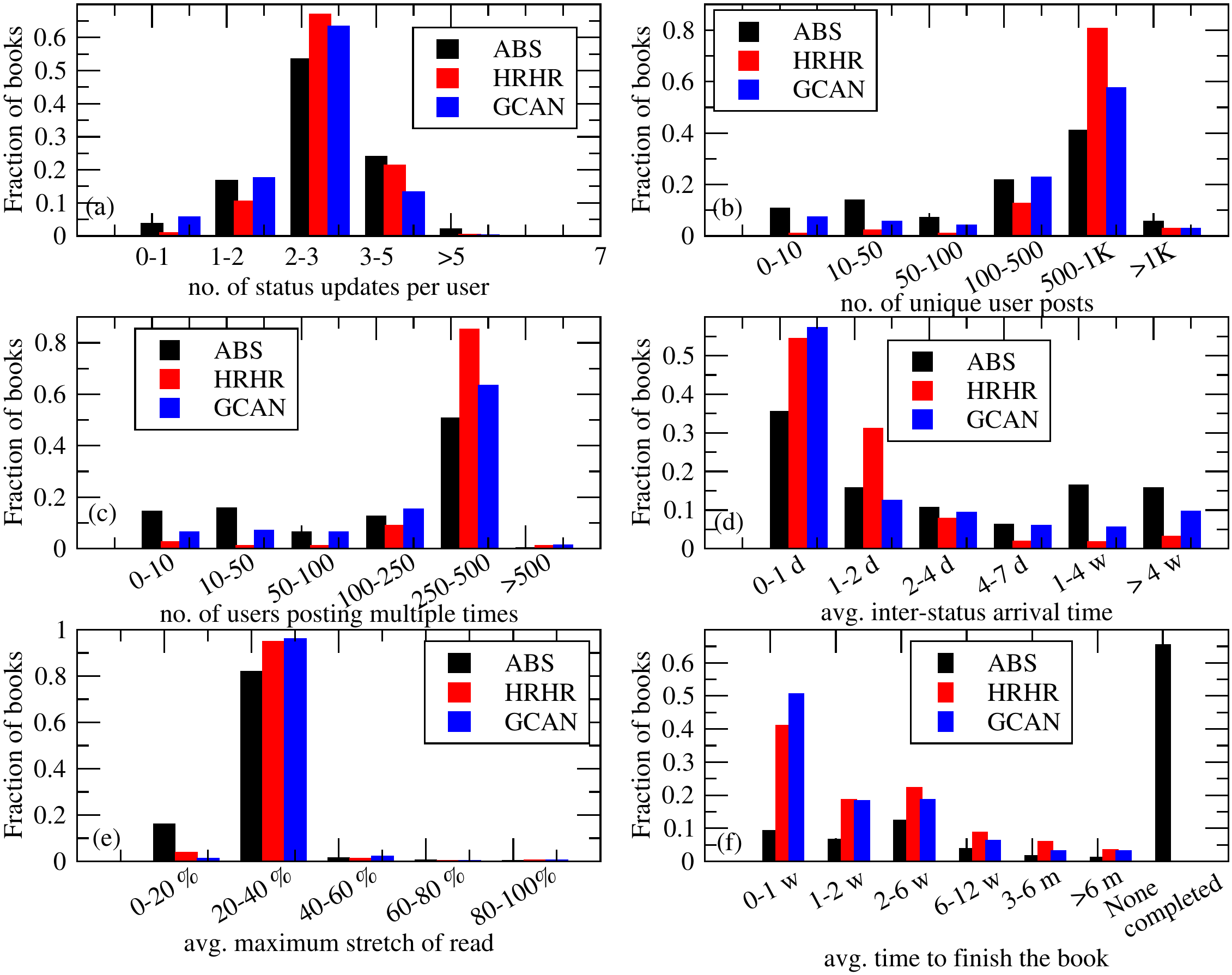}
\caption{\label{figcompstatus} \small Characteristic properties of Goodreads users' status posts: distribution of a) number of status updates per user  b) number of unique users updating status c) number of users updating multiple times d) inter-status arrival time e) average maximum stretch of reading f) average time to finish reading  for ABS vs HRHR and GCAN books.}
\end{center}
\end{figure*}

\noindent\textbf{Goodreads users' status posts}: As shown earlier, Goodreads users' reading status posting behavior is a key discriminator between the Amazon best sellers and other Goodreads books. Here we repeat the analysis in the context of the competitor books. In case of the number of status posts, we observe that the readers of the competitor books post more status updates than that of the Amazon best sellers (see fig~\ref{figcompstatus} (a)). The number of unique users posting status updates as well the users posting multiple updates are higher for the competitor books in comparison to the Amazon best sellers (see fig~\ref{figcompstatus} (b) and (c) respectively). Fig~\ref{figcompstatus} (d) shows that the average inter-status arrival time is lesser (larger fraction of books garner status posts within a day on average) for the competitor books in comparison to the Amazon best sellers. The maximum reading effort also shows distinct characteristic with the competitor book readers lying in more proportion in the 20-40\% reading zone compared to the Amazon best sellers (\ref{figcompstatus} (e)). Fig~\ref{figcompstatus} (f) shows the distribution of average time taken by the readers to finish reading the books. One notable observation is that for all the competitor books, there is at least one reader who has finished reading the book. Also the fraction of books in various time buckets are higher for the competitor books compared to the Amazon best sellers.

\begin{figure*}[h]
\begin{center}
\includegraphics[width=1\columnwidth,angle=0]{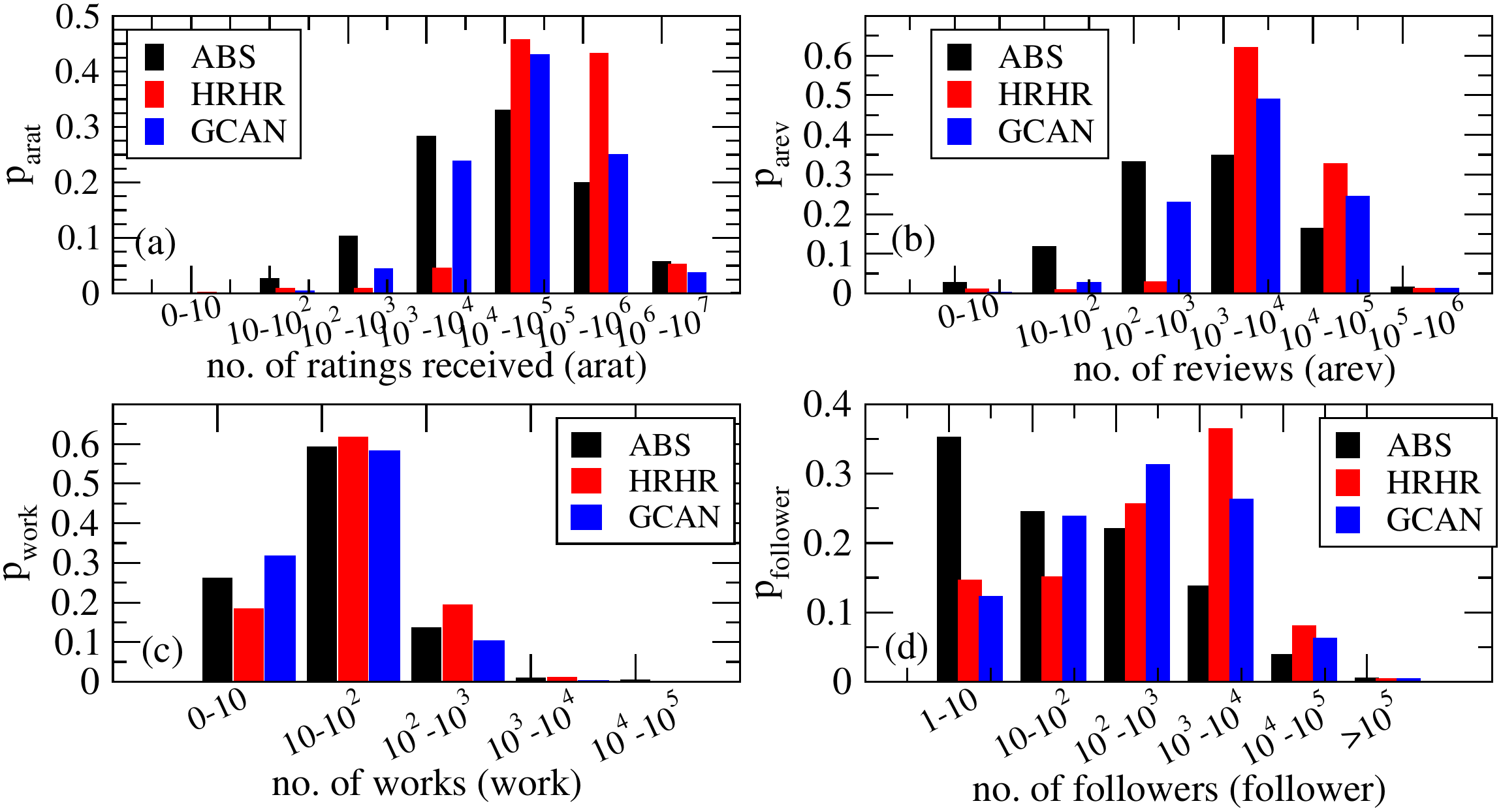}
\caption{\label{figcompauthor} \small Characteristic properties of the authors of the book: distribution of a) number of ratings received by the author b) number of reviews received c) number of literary works  d) number of followers for ABS vs HRHR and GCAN books.}
\end{center}
\end{figure*}

\noindent\textbf{Author characteristics}: Here we study the author characteristics of the competitor books. In terms of number of ratings received, the authors of the competitor books receive larger fraction of ratings in the high rating zone ($10^4-10^6$) compared to the authors of the Amazon best sellers (see fig~\ref{figcompauthor} (a)). The number of reviews distribution also follow similar trend with authors of competitor books receiving higher fraction of reviews in the high zone ($10^3-10^5$); however for the highest zone ($10^5-10^6$ reviews), the fraction is a little higher for the Amazon best sellers compared to that of the competitor books (see fig~\ref{figcompauthor} (b)). In terms of the number of literary works (see fig~\ref{figcompauthor} (c)), we observe that authors of HRHR books tend to write more than the authors of the Amazon best sellers and the GCAN books. The fan/follower distribution also shows that the authors of the competitor books have larger number of followers compared to the authors of the Amazon best sellers (see fig~\ref{figcompauthor} (d)). This analysis reveals an interesting fact that success of a book (in terms of financial profit/ best seller) does not always necessary mean best quality books. There are also books which are qualitatively better than an Amazon best seller but have not been the most successful.

We then attempt to evaluate our prediction model with these two sets of books - a) the Amazon best sellers and the HRHR books b) the Amazon best sellers and the GCAN books. For the prediction task, we consider the HRHR and GCAN books published after 2007 and take data for $t = $ 15 days, 1 month post publication to predict. Once again we compute all the features from the data available only within this time period $t$ only. We observe that the classifiers are yielding best results for ABS vs GCAN with average accuracy $\sim88\%$ (see table~\ref{tabcomp} for details). Even the prediction performance for the classifiers when ABS books are taken into competition with the HRHR books, yield average accuracy of $\sim80\%$ though the discriminating power of the classifiers are better for ABS vs GCAN classification. 

\begin{table*}[!ht]
\caption{\small Evaluation results for a) ABS vs GCAN (top 4 lines) b) ABS vs HRHR (bottom 4 lines).}
\label{tabcomp}
\begin{tabular}{ |p{2cm}|p{1.5cm}|p{1.5cm}|p{1.5cm}|p{1.5cm}|p{1.5cm}|p{1.5cm}| }
\hline
 time period & Classifier & Accuracy &Precision & Recall & F-Score & ROC Area \\ \hline
 \multirow{2}{*}{15 days}& LR &  \textbf{87.1\%}& 0.884  &   0.871  &   0.871 &   \textbf{0.898} \\\cline{2-7}
  & SVM  & 83.27\% &0.833  &   0.833  &   0.833   &   0.833 \\ \hline
 \multirow{2}{*}{1 month}&LR & \textbf{86.39\%}& 0.864 & 0.864  & 0.864 & \textbf{0.909} \\\cline{2-7}
  & SVM  & 86.1\% &0.866  & 0.861 &  0.861  &  0.861 \\ \hline

  \multirow{2}{*}{15 days}& LR &  \textbf{80.72\%}& 0.808  &   0.807 &  0.807 & \textbf{0.849}\\\cline{2-7}
  & SVM  & 79.64\%&0.796  &   0.796  &   0.796    &  0.796\\ \hline
 \multirow{2}{*}{1 month}&LR & \textbf{86.22\%}& 0.862  &   0.862   &  0.862 &  \textbf{0.919} \\\cline{2-7}
  & SVM  & 84.89\% &0.849 &    0.849 &    0.849   &   0.849 \\ \hline

  \end{tabular}
  \end{table*}
 
\section{Conclusions and future works}
In this paper, we study the characteristic properties of Amazon best sellers in terms of various Goodreads entities - books, authors, shelves and genres by analyzing a large Goodreads dataset. We observe that there exists characteristic differences between the Amazon best sellers and the other books. We then use these characteristic properties as features for a prediction model that attempts to predict whether a book will be an Amazon best seller or not.

Our proposed prediction framework achieves a very high average accuracy of {\bf 88.72}\% with high average precision and recall (\textbf{0.887}) for observation time period $t$ = 1 month. Our results also hold true for an unbalanced test data set. We observe that the user status post features are the most discriminative ones. We also evaluate our model with two different and more competitive sets of books -- HRHR and GCAN and obtain very good results (even better than the above result for ABS vs GCAN). 

There are quite a few other interesting directions that can be explored in future. One such direction could be to understand the detailed user reading dynamics focusing on various inter-dependent entities like shelves and user status posts. We are also interested in performing other cross-platform study to understand this unique dynamics between the platforms in more detail.

\bibliographystyle{splncs}
\bibliography{ref} 

\end{document}